\documentclass
{elsart5p}

\usepackage{times}

\usepackage{graphicx}

\usepackage{amsmath}

\usepackage{kbordermatrix}

\renewcommand{\vec}[1]{\boldsymbol{#1}}      
\begin{document}
\parindent 1em

\begin{frontmatter}
\vspace{-30pt}


\title{Uniaxially deformed (5,5) carbon nanotube: Structural transitions}


\author[BSU]{N.A. Poklonski\corauthref{cor}},
\corauth[cor]{Corresponding author.}
\ead{poklonski@bsu.by}
\author[BSU]{E.F. Kislyakov},
\author[BSU]{Nguyen Ngoc Hieu},
\author[BSU]{O.N. Bubel'},
\author[BSU]{S.A. Vyrko},
\author[ISAN]{A.M. Popov},
\ead{am-popov@isan.troitsk.ru}
\author[ISAN]{Yu.E. Lozovik}
\ead{lozovik@isan.troitsk.ru}

\address[BSU]{Physics Department, Belarusian State University,
Minsk 220030, Belarus}
\address[ISAN]{Institute of Spectroscopy, Troitsk 142190, Moscow Region, Russia}

\begin{abstract}
The Kekule structure of the ground state of (5,5) armchair carbon nanotube is revealed by semiempirical molecular orbital calculations. This structure has bonds with two different bond lengths, differing by 0.003~nm. The ground state has tripled (compared to undistorted case) translational period due to Peierls distortions. Two first order structural phase transitions controlled by the tension are predicted at zero temperature. These transitions correspond to 5\% and 13\% elongations of a uniaxially deformed (5,5) nanotube. The narrow gap semiconductor to metal transition is predicted at 5\% elongation of the nanotube.

\end{abstract}



\end{frontmatter}

\hyphenation{nano-tube nano-tubes}
\section{Introduction}
The studies of electronic and elastic properties of carbon nanotubes (CNTs) are actual in connection with perspectives of their applications in nanoelectronic devices~\cite{Jorio08} and in composite materials~\cite{Rul04}, and are also of fundamental interest, particularly for physics of phase transitions. For example, the superconductivity~\cite{Takesue06}, the commensurate--incommensurate phase transition in double-walled nanotubes~\cite{bichoutskaia04} and the spontaneous symmetry breaking with formation of corrugations along nanotube axis in ultrasmall CNTs (0.4~nm in diameter)~\cite{connetable05} have been considered.

Within a framework of the tight-binding model the electronic properties of CNTs are determined by their diameter and chirality~\cite{hamada92} and armchair ($n,n$) CNTs are metallic. The first clear experimental evidence of the chirality dependence of electronic properties of CNTs have been gained in STM/STS measurements~\cite{Wildoer98,Odom98}. However such kind of measurements can not resolve any bond length difference in CNTs and now~\cite{Jorio08} electronic properties are conventionally considered for CNTs with all equal graphitic C--C bond lengths.

\begin{figure*}
\hfil\includegraphics{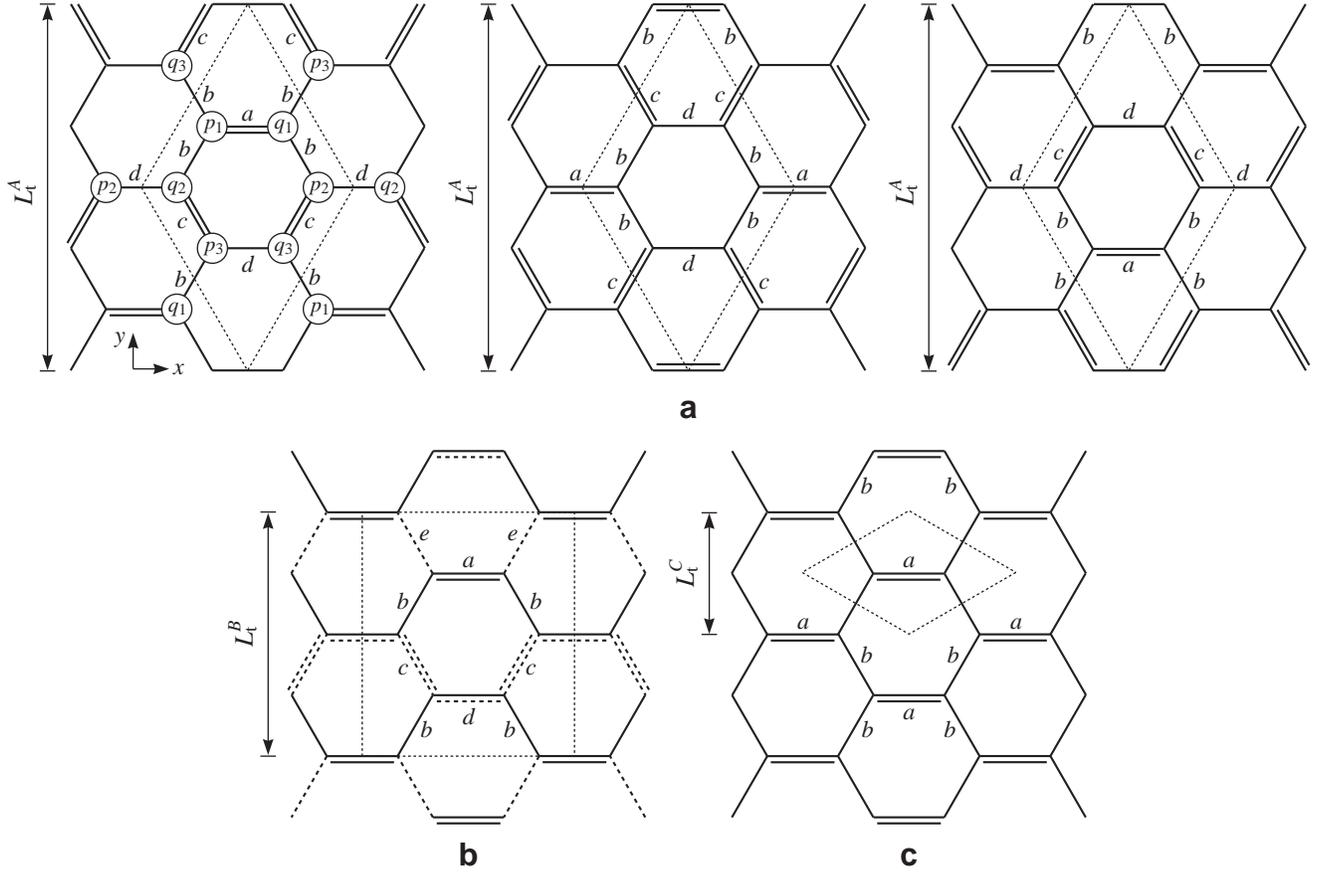} 
\caption{
Atomic structures of (5,5) carbon nanotube corresponding to: (a) three energetically equivalent structures of phase $A$ (the ground state of the undeformed nanotube); (b) phase $B$ of deformed nanotube; (c) quinoid structure of the phase $C$ of deformed nanotube and also of the transition state of the phase $A$. Dotted lines in (a), (b) and (c) are the primitive unit cells of the carbon nanotube for the phases $A$, $B$ and $C$, respectively. The coordinate system $x, y$ is used for the calculation of the Hamiltonian matrix \eqref{mat}. Translational periods $L_t^C = L_t^B/2 = L_t^A/3 = a_0\sqrt{3} = L_t$, if all bond lengths are equal to $a_0$.}\label{fig:01}
\end{figure*}

The possibility of Peierls transition in carbon nanotubes was first considered in~\cite{mintmire92}. As a result of this transition, armchair CNTs become semiconducting at low temperature and Peierls distortions lead to Kekule structure (see Fig.~\ref{fig:01}(a)) with triple translational period $L_t^A$ (three times more hexagons in translational unit cell). In early works the Peierls gap and Peierls transition temperature were estimated in the framework of Su--Schrieffer--Heeger model~\cite{Viet94,harigaya93} and by means of phenomenological electron-phonon coupling parameters~\cite{sedeki00,huang96}. For a recent review on this problem see~\cite{Piscanec07}.

By now the Kekule structure of the ground state was found only for finite
length nanotubes~\cite{Zhou04, matsuo03, nakamura03}.
Density functional theory (DFT) calculations for finite length (5,5) CNT~\cite{Zhou04,matsuo03} give 60 atom periodicity of physical properties on the length of CNT segment which is consistent with Kekule structure for infinite armchair CNTs. Moreover, X-ray crystallographic analysis of recently chemically synthesized short (5,5) CNTs~\cite{nakamura03} show the Kekule bond length alternation pattern for their geometrical structure, which is in good agreement with DFT and PM3 calculations also performed in~\cite{nakamura03}.

In the present Letter we consider the possibility of structural phase transitions connected with spontaneous symmetry breaking and controlled by uniaxial deformation of CNTs on the example of the infinite armchair (5,5) CNT. We present the PM3~\cite{Stewart89} calculations of the internal energy and geometrical structure of a uniaxially deformed infinite armchair (5,5) CNT up to 17\% elongation. For calculated geometrical structure we use simple tight-binding (H\"uckel)~\cite{Saito98} calculations (taking into account only the first nearest-neighbour interactions) to gain qualitative insight in the electronic energy band structure of a uniaxially deformed armchair (5,5) CNT.

\section{Methodology}

The MOPAC-2006 code \cite{MOPAC2006} was used in PM3 calculations of the internal energy and the structure of the (5,5) CNT. The adequacy of the PM3 parametrization of the Hamiltonian has been already demonstrated~\cite{Bubel00} by the calculation of bond lengths of the C$_{60}$ fullerene with $I_h$ symmetry: the calculated values of the bond lengths coincide with the measured ones~\cite{Leclercq93} at the level of experimental accuracy of $10^{-4}$~nm at liquid helium temperature. The PM3 parameterization gives also correct C--C bond length in graphite~\cite{Budyka05}. Comparison of different types of semiempirical calculations~\cite{Stewart07} shows that the PM3 parameterization gives the best accuracy for structure and internal energy of carbon nanostructures.

Since the Kekule structure of the infinite armchair CNTs has the triple translational period, the proper choice of the length of a computational cell along the CNT axis is necessary for adequate calculations of their geometrical structure. Namely, if one chooses the number of atoms in the computational cell of infinite ($n,n$) CNT which is not divisible on 12$n$, then the Kekule structure is excluded from consideration. This is the case in many quantum-chemical calculations of infinite armchair CNTs (see, e.g.,~\cite{Budyka05,Ogata03,Sun03}). For the best of our knowledge (see also~\cite{Zhou04}) there is only one quantum-chemical calculation~\cite{Tanaka97} of infinite armchair CNT with properly chosen computational cell for considering the Kekule structure. This early version of Complete Neglect of Differential Overlap semiempirical CNDO/2 calculation gives the quinoid type structure (see Fig.~\ref{fig:01}(c)) for the ground state of the (5,5) CNT being slightly lower in energy than the Kekule structure. On the other hand, our~\cite{Poklonski05NT297} semiempirical NDDO-type (Neglect of Diatomic Differential Overlap) calculations with the PM3~\cite{Stewart89} parameterization (as discussed above it is very well suited for the calculations of pure carbon systems) give the Kekule structure for the ground state of the (5,5) CNT.

In calculations of the ground state geometrical structure of the infinite undeformed (5,5) CNT we used computational cells with 60, 120 and 240 atoms in order to check the convergence versus the size of the computational cell. Born--von Karman periodic boundary conditions along the nanotube axis and the full geometry optimization (without any symmetry constraints) including the computational cell length were used.

\begin{figure*}
\noindent\hfil\includegraphics{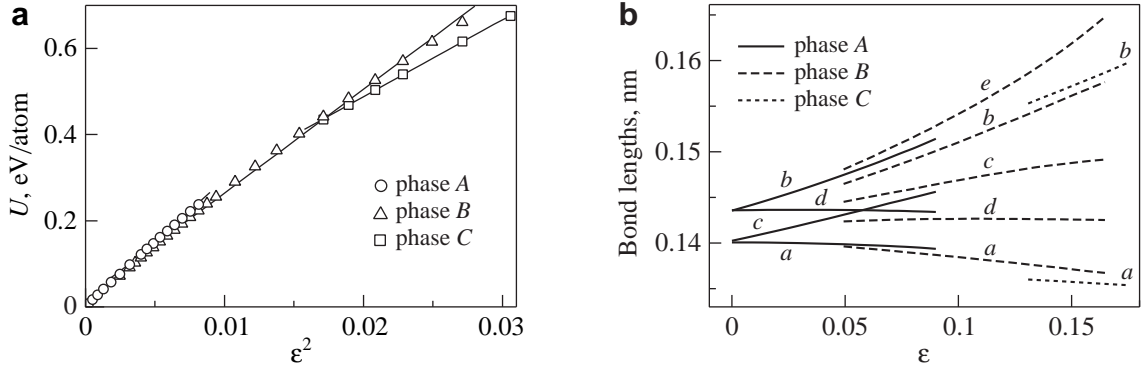} 
\caption{(a) Dependence of the internal energy $U$ on the square of elongation $\varepsilon^2$ for the (5,5) carbon nanotube, the lines are linear fits of calculated values, showing the Young's module difference for three phases. (b) Dependence of the bond lengths $a$, $b$, $c$, $d$ and $e$ on the elongation $\varepsilon$.}\label{fig:02}
\end{figure*}

\section{Ground state structure of undeformed (5,5) CNT }

The calculated Kekule structure of the undeformed (5,5) CNT ground state at zero temperature is shown in Fig.~\ref{fig:01}(a). Geometrically, this structure is possible for all armchair ($n,n$) CNTs. Three energetically equivalent atomic structures corresponding to the ground state are shown in Fig.~\ref{fig:01}(a). The transition state between these minima is shown in Fig.~\ref{fig:01}(c). The calculated energy difference between the ground (with tripled period) and transition states (with the initial period) of the (5,5) CNT is 3~meV per a carbon atom. The bond lengths have the values $a=0.1405$, $b=0.1433$, $c=0.1406$, $d=0.1434$~nm for the ground state and $a=0.1420$, $b=0.1426$~nm for the transition state. Note that the bond length difference between short ($a$ and $c$) and long bonds ($b$ and $d$) of the ground state is 6 times greater than the difference between the bonds of the transition state. As opposed to the transition state structure with all atoms lying on cylindrical surface, the Kekule structure represents the alternating rings of atoms with different radii, $R_1=0.341$~nm for rings with $p_1$ and $q_1$ atoms and $R_2=0.344$~nm for rings with $p_2$ and $q_2$ and with $p_3$ and $q_3$ atoms. Note that the same pattern (Kekule structure) with bond lengths close (with relative error $< 1\,\%$) to our calculations is revealed by DFT method for finite length hydrogen terminated (5,5) CNT C$_{160}$H$_{20}$~\cite{Zhou04}.


\section{Deformational structural phase transitions}

Since the translational period of deformed nanotube {\it a priori} is not known, we use computational cell of 120 carbon atoms to match both possibilities of tripling and doubling of the translational period of deformed CNT. The structure and internal energy of the (5,5) nanotube are calculated for different elongations (all other geometrical parameters have been optimised) $\varepsilon = (l - l_\text{eq})/l_\text{eq}$, where $l$ is the length of the computational cell, and $l_\text{eq}$ is the length of this cell, corresponding to the elongation $\varepsilon=0$. The dependences of the internal energy $U$ and the non-equivalent bond lengths with respect to $\varepsilon$ are shown in Fig.~\ref{fig:02}(a) and Fig.~\ref{fig:02}(b), respectively.

The following phases with different symmetries of the structure have been found for the uniaxially deformed (5,5) CNT: the phase $A$ with the Kekule structure and the triple translational period $L_t^A$ at small elongations $\varepsilon$, the phase $B$ with the double translational period $L_t^B$ at intermediate $\varepsilon$, and the phase $C$ with the quinoid type structure and the ordinary translational period $L_t^C$ at large $\varepsilon$. The primitive unit cells which should be used to obtain the translational unit cells of the phases $A$, $B$ and $C$ with the help of rotational (for all phases) and helical (only for the phases $A$ and $C$) symmetry transformations are shown in Fig.~\ref{fig:01}.

Since we study the system at constant length and temperature $T \to 0$, the ground state is determined by the minimum of the system internal energy. The Fig.~\ref{fig:02}(a) shows that at zero temperature the phase $A$ is the ground state of the (5,5) nanotube for the elongation $\varepsilon < 0.05$, the phase $B$ (see Fig.~\ref{fig:01}(b)) is the ground state for $0.05 < \varepsilon < 0.13$ and the phase $C$ (see Fig.~\ref{fig:01}(c)) is the ground state for $\varepsilon > 0.13$. The dependence of the bond lengths on the elongation $\varepsilon$ demonstrates that the structures of the phases are different at critical elongation values $\varepsilon_c$ where the internal energies of the phases coincide (see Fig.~\ref{fig:02}(b)). The structure change at $\varepsilon_c$ means that at zero temperature both structural phase transitions between the phases $A$ and $B$ and the phases $B$ and $C$ are first order transitions with respect to the control parameter (elongation). The Young's modules of structures are also changed at these phase transitions (see Fig.~\ref{fig:02}(a)). Fig.~\ref{fig:02}(a) shows the possibility of metastable states for the phase $A$ for elongation $0.05 < \varepsilon < 0.09$, and for the phase $B$ for elongation $\varepsilon > 0.13$. The calculations show that the phase $C$ is unstable for elongation $\varepsilon < 0.13$.

The structure without Stone-Wales (SW) defects is energetically favorable for armchair nanotubes at least for elongations $\varepsilon < 0.1$ \cite{samsonidze02}. Moreover the high barrier for formation of SW defects excludes their spontaneous formation at room temperature up to the elongation $\varepsilon = 0.15$ \cite{samsonidze02}. Thus we do not consider nanotube structures with SW defects. Note also that we consider nanotubes at elongations which are well under the fracture value ($\varepsilon=0.24$ according to AM1 calculations \cite{dumitrica03} and $\varepsilon=0.3$ according to PM3 and density functional theory-based calculations \cite{mielke04}).

Since we are considering a 1D system, the term $PV$ in the thermodynamic potentials should be replaced with the term $Fl$, where $F$ is the tension force and $l$ is the length of the nanotube (a positive force value corresponds to contraction). The chemical potential $\mu_i$ for any structural phase $i$ of the nanotube takes the form
\begin{equation*}
   \mu_i=u_i-Ts_i+FL_i,
\end{equation*}
where $u_i$, $s_i$ and $L_i$ are the internal energy, the entropy and the length per one particle of the 1D system. It is well known that in 1D systems at temperature $T \neq 0$ a co-existence of phases with a boundary between them is not possible \cite{Landau}. The phase transitions in 1D systems have a crossover character, and the crossover region (with respect to the control parameter) increases with temperature, see Refs.~\cite{Kagoshima88, Zaitsev-Zotov04}. In the crossover region the nucleus of another phase is formed as a result of thermal fluctuations. From the equality of the chemical potentials ($\mu_i = \mu_j$) of two phases $i$ and $j$ the line of the critical values of the tension force $F_c$ and temperature $T_c$, which correspond to the middle of the crossover region on the tension force-temperature ``phase diagram'' of the system in equilibrium, can be obtained
\begin{equation*}
   F_c=\frac{u_i-u_j+T_c(s_i-s_j)}{L_i-L_j}.
\end{equation*}
These structural ``phase transitions'' in nanotubes can be determined by peculiarities of the dependences of elastic constants or of the specific heat on temperature or on the tension force.

The dependence of the internal energy $U$ on the length $l$ of the computational cell is interpolated using the Hooke's law
\begin{equation*}
   U=U_0+\frac{\nu (l-l_\text{eq})^2}{2}=U_0+\frac{\nu \varepsilon^2l^2_\text{eq}}{2},
\end{equation*}
where $\nu$ is the coefficient of elasticity of a CNT with the length~$l$.

The Young's modulus of the (5,5) CNT at the elongation $\varepsilon=0$ has the following form
\begin{equation}\label{young}
  Y = \frac{\nu l_\text{eq}}{2\pi Rw},
\end{equation}
where $R=(R_1+R_2)/2$ is the average radius of the CNT corresponding to the elongation $\varepsilon=0$, $w=0.34$~nm is the effective thickness of the wall \cite{mielke04,vanlier00,sanchez-portal99,bichoutskaia06}. As a result the value $Y \approx 1.2$~TPa is calculated. The calculated Young's modulus of the (5,5) CNT with the Kekule structure is in good agreement with the values calculated for the (5,5) CNT with the quinoid type structure: 1.1~TPa by PM3 calculations \cite{mielke04}, 1.06~TPa by Hartree--Fock method \cite{vanlier00}, 0.95~TPa \cite{sanchez-portal99}, 0.96~TPa \cite{mielke04} and 1.03~TPa \cite{bichoutskaia06} by density functional theory-based calculations \cite{bichoutskaia06}.

Since the radius $R$ depends on the elongation $\varepsilon$, it is not convenient to use Eq.~\eqref{young} for expanded nanotubes. Moreover, the coefficient of elasticity $\nu$ is the quantity which can be measured experimentally. Thus, we have calculated the changes of the coefficient of elasticity at the structural phase transitions: $\nu_B=0.82\nu_A$ and $\nu_C=0.74\nu_B$, where $\nu_A$, $\nu_B$ and $\nu_C$ are the coefficients of elasticity for the phases $A$, $B$ and $C$, respectively.

\section{Electronic structure of a deformed nanotube}

\begin{figure}[!t]
\noindent\hfil\includegraphics{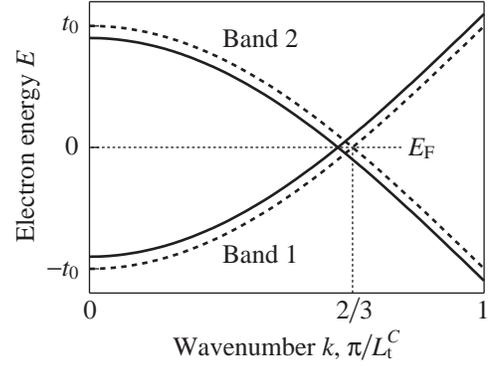}
\caption{Half-filled bands 1 and 2 intersecting at the Fermi level ($E = 0$) of quinoid type armchair carbon nanotubes (solid lines) for the case $t_a = 1.1t_b$, $t_b = t_0$. The dotted lines $E_{1,2}(k)$ are for equal lengths of bonds ($a = b$) and $t_a = t_b = t_0$.}\label{fig:03}
\end{figure}

In the tight-binding approximation~\cite{Saito98}, the band structure of armchair CNTs with all equal bond lengths has two half-filled bands intersecting at the Fermi level $E_\text{F}$. They are shown by dotted lines in Fig.~\ref{fig:03}.

In the case of the quinoid type structure of an armchair CNT (Fig.~\ref{fig:01}(c)) these bands can be expressed as (solid lines in Fig.~\ref{fig:03})~\cite{Okahara94}:
\begin{equation*}
   E_{1,2}(k) = \pm t_a[1 - 2(t_b/t_a)\cos(kL_t^C/2)],
\end{equation*}
where $t_a$ and $t_b$ are the resonance integrals for bonds $a$ and $b$, respectively, $k$ is the wavenumber of an electron along the CNT axis, and $L_t^C$ is the translational period. In this case the armchair CNT remains metallic.

The calculation of the band structure of an armchair CNT with four non-equal bonds (corresponding to the Kekule structure of the phase $A$) is reduced in the tight-binding approximation to the diagonalization of the sixth order Hamiltonian matrix~\cite{Viet94} and the standard zone-folding procedure~\cite{Saito98}. The standard procedure implies that all atoms are placed in a plane. For the primitive unit cell in Fig.~\ref{fig:01}(a) the Hamiltonian matrix has the form:
\begin{equation}\label{mat}
\renewcommand{\arraystretch}{1}
   \kbordermatrix{ & p_1 & p_2 & p_3 & q_1 & q_2 & q_3 \\
p_1&0&0&0&t_a\mathrm{e}^{-i\vec{k}\vec{r}_1}&t_b\mathrm{e}^{-i\vec{k}\vec{r}_2}&t_b\mathrm{e}^{-i\vec{k}\vec{r}_3}\\
p_2&0&0&0&t_b\mathrm{e}^{-i\vec{k}\vec{r}_3}&t_d\mathrm{e}^{-i\vec{k}\vec{r}_4}&t_c\mathrm{e}^{-i\vec{k}\vec{r}_5}\\
p_3&0&0&0&t_b\mathrm{e}^{-i\vec{k}\vec{r}_2}&t_c\mathrm{e}^{-i\vec{k}\vec{r}_6}&t_d\mathrm{e}^{-i\vec{k}\vec{r}_4}\\
q_1&t_a\mathrm{e}^{i\vec{k}\vec{r}_1}&t_b\mathrm{e}^{i\vec{k}\vec{r}_3}&t_b\mathrm{e}^{i\vec{k}\vec{r}_2}&0&0&0\\
q_2&t_b\mathrm{e}^{i\vec{k}\vec{r}_2}&t_d\mathrm{e}^{i\vec{k}\vec{r}_4}&t_c\mathrm{e}^{i\vec{k}\vec{r}_6}&0&0&0\\
q_3&t_b\mathrm{e}^{i\vec{k}\vec{r}_3}&t_c\mathrm{e}^{i\vec{k}\vec{r}_5}&t_d\mathrm{e}^{i\vec{k}\vec{r}_4}&0&0&0},
\end{equation}
where $t_a$, $t_b$, $t_c$, $t_d$ are the resonance integrals corresponding to the bonds $a$, $b$, $c$, $d$, respectively; $\vec{k} = k_x\vec{e}_x + k_y\vec{e}_y$ is the electron wavevector; $\vec{e}_x$, $\vec{e}_y$ are the unit vectors of the coordinate system; $\vec{r}_1 = p_1q_1 = a\vec{e}_x$, $\vec{r}_2 = p_1q_2 = p_3q_1 = -b_x\vec{e}_x - b_y\vec{e}_y$, $\vec{r}_3 = p_1q_3 = p_2q_1 = -b_x\vec{e}_x + b_y\vec{e}_y$, $\vec{r}_4 = p_3q_3 = p_2q_2 = d\vec{e}_x$, $\vec{r}_5 = p_2q_3 = -c_x\vec{e}_x - c_y\vec{e}_y$, $\vec{r}_6 = p_3q_2 = -c_x\vec{e}_x + c_y\vec{e}_y$ are the vectors of bonds between atoms. For the given (from the PM3 calculations) bond lengths $a$, $b$, $c$, $d$ and the translational period $L_t^A$ we numerically diagonalized~\eqref{mat} to obtain the electron energy dispersion relation $E(\vec{k})$. The condition for circumferential quantization $k_x = 2\pi m/C_h$, $m = -4, \ldots, 5$, where $C_h = 5(2c_x + 2d)$ is the circumference length (chiral vector module) of the (5,5) CNT, gives electronic energy bands $E_m(k_y)$ for the nanotube. We assume that the dependence of the resonance integral on the C--C bond length $a_\text{C--C}$ is~\cite{Harrison89}:
\begin{equation*}
   t = (a_0/a_\text{C--C})^2t_0,
\end{equation*}
where $a_0 = 0.142$~nm, and $t_0 = 2.6$~eV are parameters for the bond length and resonance integral~\cite{Wildoer98,Odom98} of the CNT with all equal bond lengths.

\begin{figure*}
\noindent\hfil\includegraphics{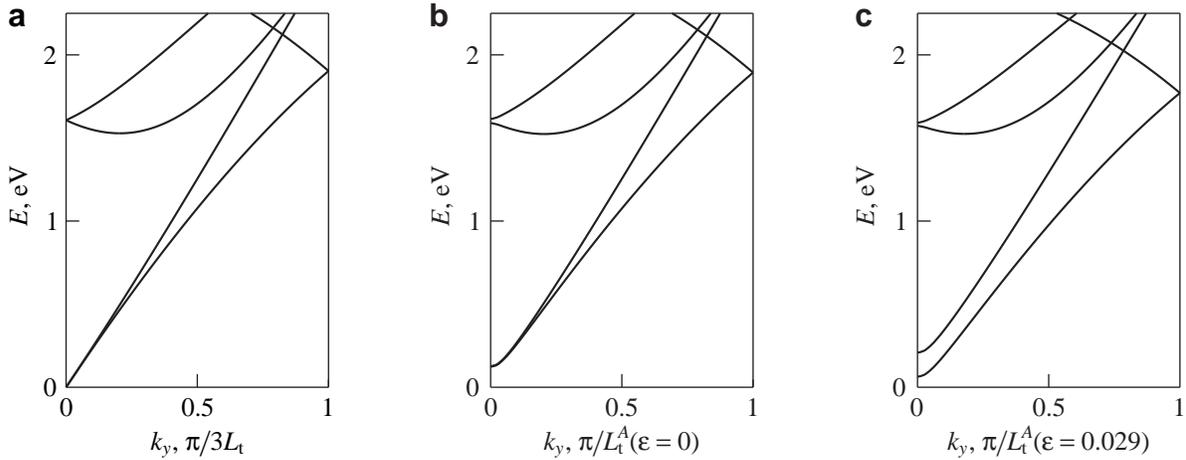}
\caption{Band structure of the (5,5) carbon nanotube near Fermi level ($E = 0$): (a) undistorted nanotube (all equal bonds) with artificially tripled period, (b) for the case of Kekule ground state distortion (Fig.~\ref{fig:01}(a)) with $a = c = 0.140$~nm and $b = d = 0.143$~nm, and (c) for the phase $A$ at the elongation $\varepsilon = 0.029$. Only conduction bands are shown, the valence bands are the same with the opposite sign.}\label{fig:04}
\end{figure*}

For the undistorted nanotube (all equal bonds) with the artificially tripled along nanotube axis translational period, the electronic energy band structure in the mapped Brillouin zone is shown in Fig.~\ref{fig:04}(a) for the conduction band near Fermi level. In the framework of our approximations (the overlap integral for the nearest neighbour atoms is neglected) the valence band is the same with the opposite sign. In this case the intersection point of half-filled bands is folded to the $\Gamma$ point of the Brillouin zone. For the distortions shown in Fig.~\ref{fig:01}(a) the energy (Peierls-type) bandgap appears between the valence and the conduction bands of the nanotube (see Fig.~\ref{fig:04}(b)). The value of this gap is found to be $E_g = 0.24$~eV. STM measurements of the electron density of states show $E_g = 0.11$~eV for the armchair (7,7) CNT at liquid helium temperature \cite{ouyang01}. We believe that the observed bandgap can be explained by the Peierls transition.

The diagonalization of matrix \eqref{mat} for $\varepsilon = 0.029$ with the PM3 calculated bond lengths of the (5,5) CNT and zone-folding gives the electronic energy band structure shown in Fig.~\ref{fig:04}(c). As a result of deformation the lowest conduction band with $m = 0$ is splitted at $k_y = 0$. The diagonalization of \eqref{mat} for $\vec{k} = 0$ gives for the bandgap $E_g = 2(t_d - t_c)$, i.e., almost linear dependence on $\varepsilon$ according to Fig.~\ref{fig:02}(b). Note that the analogous formula for the bandgap is obtained for the case of \emph{trans}-polyacetylene~\cite{Saito98}.

\begin{figure}[!h]
\noindent\hfil\includegraphics{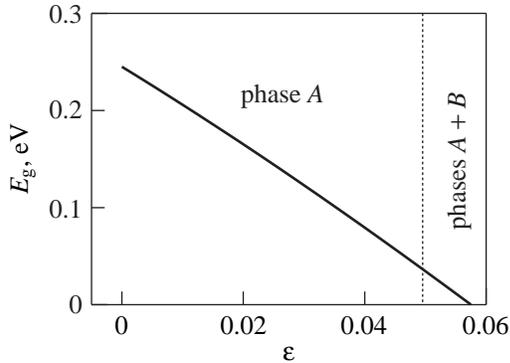}
\caption{Calculated dependence of the (5,5) carbon nanotube energy bandgap $E_g$ (at wavevector $\vec{k} = 0$) for the phase $A$ on the elongation $\varepsilon$. The vertical dotted line indicates the structural phase transition.}\label{fig:05}
\end{figure}

The results of the bandgap $E_g$ calculations for different elongations $\varepsilon$ of the phase $A$ are presented in Fig.~\ref{fig:05}. At the elongation $\varepsilon = 0.05$ the structural phase transition takes place, and the bandgap becomes zero for the phase $B$ due to the double translational period $L_t^B$ of this phase (Fig.~\ref{fig:01}(b)).

\section{Conclusion}
The atomic structure of the (5,5) carbon nanotube is calculated for its quasistatic expansion up to the 17\% elongation. Two first order structural phase transitions with the change of the atomic structure symmetry are revealed at the nanotube expansion. It is shown that for the elongation greater than 5\% the structure of the nanotube corresponds at any temperature to the metallic phases without the Peierls gap in the electron spectrum. The narrow gap semiconductor to metal phase transition at nanotube elongation can be used for elaboration of nanotube-based stress nanosensors.

\section*{Acknowledgments}
This work has been partially supported by the BFBR (grant No. F08R-061) and RFBR (AMP and YEL grants 08-02-00685 and 08-02-90049-Bel).



\end{document}